\documentclass[conference]{IEEEtran}

\usepackage{cite}

\usepackage[pdftex]{graphicx}
\graphicspath{{figures/}}
\DeclareGraphicsExtensions{.png}

\usepackage{algorithm}
\usepackage{algpseudocode}
\algnewcommand{\LineComment}[1]{\State \(\triangleright\) #1}

\usepackage{url}

\usepackage[caption=false,font=normalsize,labelfont=sf,textfont=sf]{subfig}

\usepackage[]{fancyhdr} %
\newcommand{\changefont}{\fontsize{9}{9}\selectfont}
\IEEEoverridecommandlockouts
\begin{document}

\title{Power Hardware-In-the-Loop Testing of a \\ Peer-to-Peer Energy Trading Framework}

\author{\IEEEauthorblockN{Thomas Perrau, Maksim Stojkovic and Gregor Verbi\v{c}}
\IEEEauthorblockA{School of Electrical and Information Engineering\\The University of Sydney\\Sydney, Australia}}

\maketitle
\thispagestyle{fancy}
\pagestyle{fancy}

\begin{abstract}
This paper demonstrates the value of power hardware-in-the-loop (PHIL) testing for the study of peer-to-peer (P2P) energy trading. P2P has emerged as a promising candidate for coordinating large numbers of distributed energy resources (DER) that pose a risk to network operations if left unmanaged. The existing literature has so far relied on pure software simulations to study DER and distribution networks within this context. This requires the development of simplified models for complex components due to the computational limitations involved. Issues that arise through the operation of physical hardware in real-world applications are therefore neglected. We present PHIL testing as a solution to this problem by exhibiting its ability to capture the complex behaviors of physical DER devices. A high-fidelity PHIL test environment is introduced that combines key hardware elements with a simulated network model to study a P2P trading scenario. The initial findings reveal several underlying challenges of coordinating DER that are not typically discussed in prior works.
\end{abstract}


\section{Introduction}

Distributed energy resources (DER) are quickly becoming a critical component of modern electricity networks. The Australian Electricity Market Operator (AEMO) projects that the installed capacity of DER will double or triple by 2040 and form between 13\% to 22\% of all electricity generation in the National Electricity Market \cite{AEMO}. This level of growth is also being observed in many other jurisdictions around the world \cite{Newport}.

High penetrations of DER lead to violations of network constraints when not managed effectively \cite{GuerreroTransactive, Newport, HighPVImpacts}. This has sparked research into methods of coordinating large numbers of these devices that preserve system integrity and incentivize efficient usage. Such methods aim to strike a balance between maximizing social welfare and minimizing network instability.

There are several leading approaches towards solving this emerging problem. Among these is peer-to-peer (P2P) energy trading, which employs a market-driven strategy to encourage efficient DER operation. Under this strategy, energy can be transacted directly between participating energy \textit{prosumers} and traditional consumers. Excess generation from DER can be routinely auctioned to potential buyers who wish to purchase energy at lower prices than the retail value. Buyers and sellers are matched during predetermined trading periods, with transacted energy then being dispatched to fulfill the conditions of each executed trade.

This structure brings added efficiency to several key areas of electricity networks. Local supply-demand balancing is encouraged through the market clearing process which reduces overall system losses \cite{GuerreroTransactive}. P2P trading reliably converges to optimal pricing in each trading window which has a positive effect on social welfare \cite{DecentralP2P, DistributedP2PBlockchain}. Other benefits include the improved utilization of excess generation (or reduced energy \textit{spillage}) and clear incentives for new DER installations \cite{NetAwareP2P}. 

A key limitation of current P2P trading frameworks is the exclusion of network constraints. Trading without oversight from a network operator can lead to violations of voltage, frequency and thermal operating limits \cite{DecentralP2P}. Methods of intervention by an entity such as a Distribution System Operator (DSO) have subsequently been proposed in \cite{NetAwareP2P} to address this.

A review of existing literature has revealed that the testing of P2P DER coordination methods has largely been confined to simulations and real-world trials. While such approaches do yield valuable insight into overall performance, each comes with distinct limitations. Pure software simulations allow for a wide range of operating scenarios to be tested but often lack the ability to model complex system elements accurately. In contrast, real-world trials provide the most accurate test setting but at the cost of flexibility to model a range of potential operating scenarios. 

Marrying the two approaches - explicit hardware device testing and software simulations - can yield the benefits of both testing methods \cite{EuroWhitebookHIL}. This has led to the development of hybrid Power Hardware-In-the-Loop (PHIL) testing techniques which have garnered significant attention in this research space. The complex behavior exhibited by DER devices presents a challenge that is ideal for the application of this technology.

This observation motivated us to apply state-of-the-art PHIL techniques to the study of P2P energy trading. A full hardware prosumer model - containing local PV generation, battery storage and variable load - was interfaced through a Spitzenberger \& Spies linear power amplifier to a RTDS real-time digital simulator running a simple distribution network model. Equipment was selected to maximize the accuracy of operating data transferred between the software and hardware test components. A wireless communication network was also implemented through the use of Raspberry Pi devices to replicate real-world data collection conditions.

This paper aims to demonstrate the efficacy of P2P trading for coordinating multiple prosumers in a high-fidelity environment. Specific attention is given to the effects of hardware devices on the overall performance of the scheme.

The remainder of this paper is structured as follows. In Section \ref{P2P}, a detailed description of the P2P energy trading framework is provided. Section \ref{PHIL} summarizes the challenges and design considerations of PHIL test environments. Section \ref{usydPHIL} then gives an overview of the PHIL test setting used to study P2P energy trading. Preliminary results are then presented in Section \ref{results} to demonstrate the performance of the P2P trading approach on a simple radial distribution system. Concluding remarks are provided in Section \ref{conclusion}.

\section{P2P Energy Trading}
\label{P2P}

P2P energy trading differs from traditional electricity market structures by allowing prosumers to trade energy directly with neighboring users \cite{NetAwareP2P}. This has been found to introduce many efficiencies to both the transactive environment and physical generation of electricity \cite{GuerreroTransactive, FrameworkLocalTrading, P2PFuture, GameTheoryP2P, DecentralP2P, StudyEnergyTrading, NetAwareP2P, Blockchain}. In particular, P2P facilitates:

\begin{itemize}
	\item improved incentives for prosumers compared to traditional retail arrangements,
	\item reduced spillage of DER generation due to improved prosumer coordination,
	\item lower grid losses by encouraging shorter transmission distances between consumers and generation sources, and
	\item less reliance on intermediaries to perform transactions.
\end{itemize}

These benefits can be largely attributed to a core outcome of P2P energy trading: the transition from a \textit{centralized} architecture to a \textit{decentralized} architecture. While this in particular applies to the way electricity is generated and transacted, it also extends to other components of electricity networks. Recent work has identified the need for decentralized control, computation and communication infrastructure to overcome issues of scalability when coordinating large numbers of DER \cite{NetAwareP2P, FrameworkLocalTrading, CoopPEVs}. 

From a market perspective, there have been several candidate methods proposed in the literature for hosting P2P energy trading. These largely differ in the overall market structure used and the market operating strategy employed. Market structures range from discrete-time call auctions to continuous double auctions (CDAs), with each method offering different levels of scalability, awareness of network constraints, and computational overheads \cite{FrameworkLocalTrading}. Operating strategies range from centralized architectures that employ cloud-based infrastructures to decentralized approaches that use distributed ledger technologies (DLTs) such as blockchains \cite{P2PSmartHomes}.

\begin{figure}[!t]
	\centering
	\includegraphics[width=2.5in]{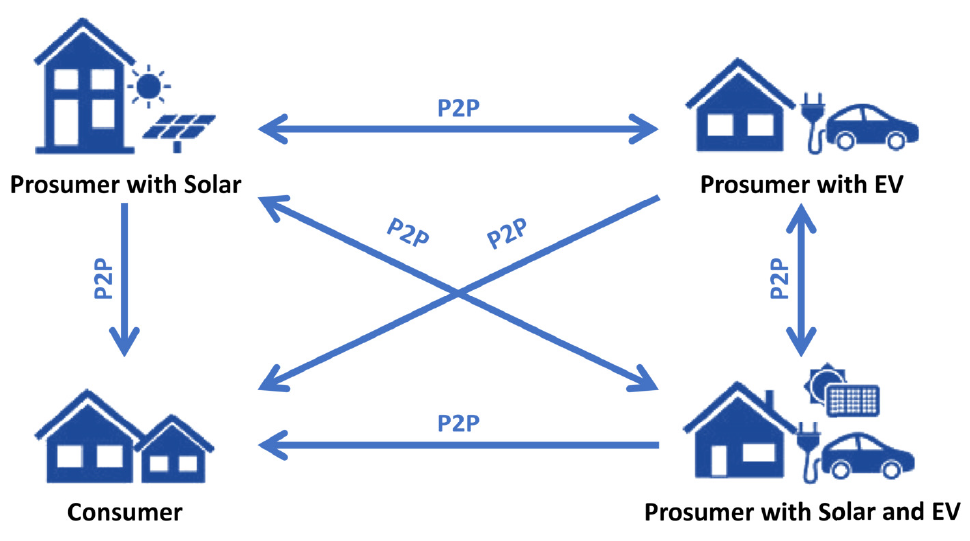}
	\caption{Illustration of P2P energy trading among participating prosumers \cite{P2PFuture}.}
	\label{P2P_Illustration}
\end{figure}

The use of a CDA has gained particular attention for being the chosen format for stock and commodity markets around the globe \cite{ZIPEfficiency, DoubleAuctions, PredictivePowerZIP}. In addition, CDAs composed of rational participants are Pareto-improving and result in a highly efficient allocation of resources \cite{NetAwareP2P, ZIPEfficiency}. This makes them an attractive option for optimizing DER usage in a P2P trading environment.

Under a CDA P2P format, buyers and sellers of energy are matched according to a set of pre-defined trading rules. Participants lodge either buy orders (\textit{bids}) or sell orders (\textit{asks}) for electricity at any time during each trading period. Sellers of energy are solely composed of prosumers with dispatchable DER. Priority is always given to the best standing offer (highest bid or lowest ask), with matches occurring when the best bid exceeds the best ask at any given time \cite{DoubleAuctions}. If no match is found for a new order, it is stored until either a match is found or the current trade period ends, in which case, all outstanding orders are canceled. All executed trades are settled immediately after each trading interval, with sellers dispatching energy according to each contract. Any energy not secured in the P2P market to satisfy a trader's demand is typically purchased through a traditional retailer.

Market participants are frequently modeled as \textit{Zero-Intelligence Plus} (ZIP) traders to represent human decision-making in an auction setting \cite{DoubleAuctions, PredictivePowerZIP}. This assumes that orders are submitted randomly by participants without any detailed underlying strategy. Bid and ask prices are instead adjusted by each trader based on the most recently matched orders. For example, traders with a bid price lower than the last executed order will decrease their margins by raising the previous bid amount.

Despite demonstrating excellent performance for improving social welfare in energy markets, several drawbacks of P2P energy trading schemes have been identified. Foremost among these is the absence of network constraint considerations and the spillage of DER generation that is not traded in the P2P market \cite{GuerreroTransactive}. Current research is being conducted into overcoming these challenges and transforming theoretical P2P approaches into real-world solutions.

\section{PHIL Testing}
\label{PHIL}

PHIL testing seeks to gain the benefits of pure software simulations and real-world hardware experimentation by combining elements from both methods \cite{EuroWhitebookHIL}. Traditional pure-software simulations offer a high degree of flexibility for power system studies by allowing the analysis of a wide range of scenarios. However, this comes at the cost of accuracy due to the need to develop software models for highly-complex hardware devices. In contrast, real-world testing offers the reverse; the presence of physical hardware yields high accuracy but lowers testing flexibility due to the risks of damaging equipment and endangering personnel. PHIL testing embeds complex hardware devices within a software environment capable of performing simulations in real time, resulting in both a highly accurate and flexible test environment.

Interest in PHIL testing methods has been increasing within the field of power systems engineering \cite{EuroWhitebookHIL, ExpLabFieldDemos, SurvRTSHILActivities, MultiphysicsTestBed, UniPlatformHILTesting, PHILHolisticApproach, PHILforGeninLVSystems, PHILforDG, AccuracyofPHIL, PHILStabilityRealPV, DistPHILGridFormConvInterface}. Technological advancements in Real-Time Simulation (RTS) and power interfacing devices have improved the fidelity of PHIL test environments for studying the complex dynamic behavior of electricity networks \cite{AdvancesRTS, PowerInterfacesPHIL}. This has led to recent applications of PHIL techniques for the study of DER integration due to the complexity of the hardware interactions involved. 

The two primary challenges of PHIL testing are the preservation of system stability and simulation accuracy \cite{EuroWhitebookHIL, PHILforGeninLVSystems, PHILforDG, PHILCapabilities, PHILStabilityRealPV, PowerInterfacesPHIL, ImpactTimeDelaysPHIL, TimeDelaysPHIL, StabilizationPHIL}. As PHIL consists of high-power signals, instability can quickly lead to equipment damage and the endangerment of personnel \cite{PHILCapabilities}. The ideal test environment should therefore be capable of capturing system behavior with high accuracy while avoiding unbounded growth in any of the signals involved.

This is difficult in practice due to the inherent errors introduced by the equipment used. In particular, the interface between the hardware and software components of the test system has a significant impact on overall stability and accuracy \cite{PHILStabilityRealPV}. Errors are introduced through sensor noise, time delays, finite signal sampling rates and the response characteristics of the power amplifier device \cite{PHILCapabilities}. These culminate in lower quality test results and an increased risk of system collapse when amplified through the power interface device.

Recent work has been done on quantifying the effects of individual interface elements on overall PHIL stability \cite{PowerInterfacesPHIL, ImpactTimeDelaysPHIL, TimeDelaysPHIL}. This has allowed the time delays introduced by different power amplifiers, communication methods, RTS devices and other components to be directly contrasted when making design decisions. Improved stability and accuracy analysis are gained, which can be used to establish concrete operating boundaries for a particular PHIL test environment.  

Simulation errors can be mitigated through the use of more advanced equipment. However, this comes with additional cost and may not always be possible within the budget of a development team. The trade-off between cost and system accuracy is, therefore, one of the major considerations in the field of PHIL testing.

\begin{figure}[!t]
	\centering
	\includegraphics[width=\columnwidth, keepaspectratio]{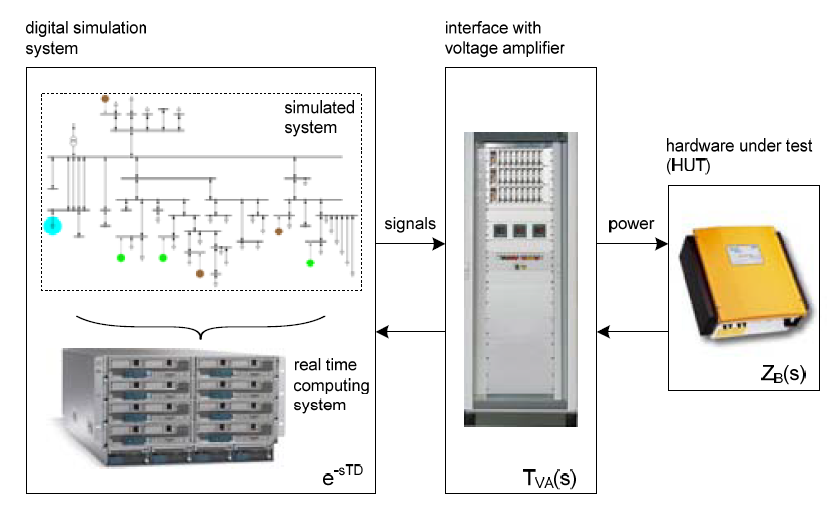}
	\caption{Typical arrangement of a PHIL test system \cite{PHILforDG}.}
	\label{PHIL_Setup}
\end{figure}

\begin{figure}[!t]
	\centering
	\includegraphics[width=\columnwidth, keepaspectratio]{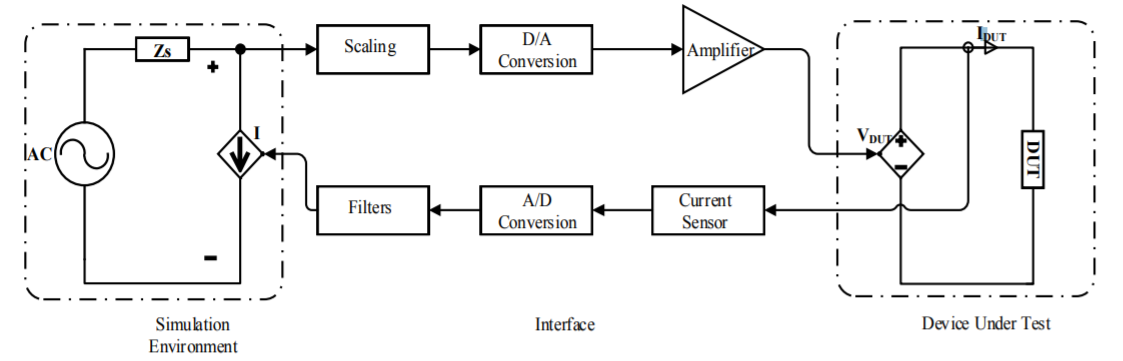}
	\caption{Basic signal block diagram of a PHIL interface using the Ideal Transformer model \cite{PHILStabilityRealPV}.}
	\label{PHIL_Interface}
\end{figure}

\section{PHIL Testing Set-Up}
\label{usydPHIL}

The PHIL test environment at the University of Sydney was designed to study DER coordination methods. The general concept was to model a typical residential prosumer entirely in hardware for connection to a real-time simulated distribution network model. This would allow all complex hardware interactions to be captured explicitly within the simulation environment. Their effects on the wider network could then be observed under different operating scenarios.

The typical prosumer was assumed to possess on-roof PV generation, local energy storage, and a variable load. This was realized in hardware using the following key pieces of equipment:

\begin{itemize}
	\item 1 × Spitzenberger \& Spies APS5000 Four-Quadrant Amplifier
	\item 1 × SMA Sunny Boy Solar Inverter
	\item 1 x Spitzenberger \& Spies PVS3000 Photovoltaic Simulator
	\item 1 × SMA Sunny Island Battery Inverter
	\item 1 × LG Chem RESU10H Lithium-Ion Battery
	\item 1 × Chroma 63803 Electronic Load Bank
\end{itemize}

The communications required between the market platform\footnote{The market platform serves the sole purpose of matching buyers and sellers. This role can be viewed as that of an auctioneer in a local energy market. It is oblivious to any overarching market structures, and so can be considered truly decentralized.} and each prosumer were also modeled explicitly in hardware. To do this, four other neighboring prosumers were simulated on Raspberry Pi devices and interfaced wirelessly to a central computer hosting the CDA trading platform. This allowed the effects of communication delays and data loss on system operations to be directly incorporated.

An RTDS Technologies Mid-Size Cubicle was used to implement the simulated distribution model. The model design was done using the corresponding RSCAD graphical user interface. The real-time monitoring and control of network simulations were also done through this software.

\begin{figure*}[!t]
	\centering
	\includegraphics[width=\textwidth, keepaspectratio]{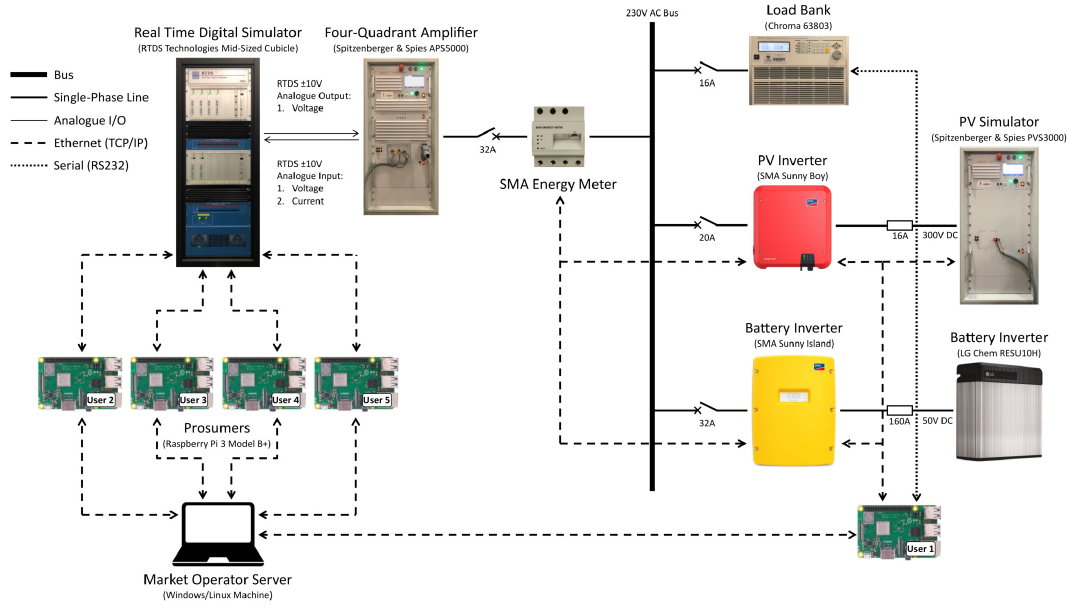}
	\caption{University of Sydney PHIL test environment design.}
	\label{test_bed_overview}
\end{figure*}

The general arrangement of the PHIL test environment is shown in Figure \ref{test_bed_overview}.

A key goal of this test environment was to obtain a very high level of accuracy for observing prosumer behavior. As a result, the interface between the software and hardware components were made to minimize time delays and noise as much as possible. 

A linear power amplifier was chosen to achieve higher dynamic performance than switched-mode and synchronous generator amplifiers seen in other PHIL installations \cite{PowerInterfacesPHIL}. This amplifier operates with a slew rate of 52V/$\mu$s, which corresponds to time delays under 5$\mu$s for most operations \cite{SSPowerAmplifier}. In contrast, the switched-mode amplifier used in \cite{TimeDelaysPHIL} was found to introduce a time delay of 66.67$\mu$s. Improved system accuracy and stability was gained as a result.

The analogue connections between the RTDS and power amplifier were replaced with a fiber-optic Aurora link after initial testing. This is known to be faster, more efficient and does not require anti-aliasing filters when compared to analogue interfaces \cite{TimeDelaysPHIL}. The work in \cite{ImpactTimeDelaysPHIL} found a time delay of $<$1$\mu$s for the Aurora protocol and 20$\mu$s for a standard analogue connection.

The installation of the Aurora link was done in response to stability and accuracy issues uncovered during the initial testing. The results in Section \ref{results} were therefore obtained using analogue communications, with a discussion on the potential impact presented in Section \ref{PHIL_behaviour}

\section{PHIL Testing Results}
\label{results}

Preliminary testing of a basic P2P market framework was conducted on the PHIL test environment to verify its performance. Tests were done prior to the installation of the Aurora link using a standard analogue interface as shown in Figure \ref{test_bed_overview}.

The P2P market operating algorithm was modeled off the approach developed in \cite{NetAwareP2P}. A CDA with hourly trading periods was used as the market structure and market participants were assumed to be ZIP traders. The matching of orders by the market platform was simulated using Algorithm \ref{alg1}. Participating ZIP energy traders updated their asks and bids in response to market events using Algorithm \ref{alg2}. 

Any energy not secured through the P2P market to satisfy demand was purchased at the time-of-use (ToU) rates of a typical retailer. Likewise, surplus PV generation not able to be sold in the market was instead settled at a standard retail feed-in-tariff (FiT). The pricing used in the simulation was adapted from the 2021 residential standing offer of a large Australian energy retailer (AGL) \cite{AGLOffer}.

\begin{algorithm}
\caption{CDA Order Matching Process}
\label{alg1}
	\begin{algorithmic}[1]
		\While{market is open during time slot}
			\State randomly select a new ZIP trader 
			\LineComment{Process orders with given (price, quantity, time)}
			\If{trader is buyer}
				\State add new order $o_b(p_b, q_b, t_b)$ to order book
			\Else
				\State add new order $o_s(p_s, q_s, t_s)$ to order book
			\EndIf
		
			\LineComment{Match orders based on price-time priority}
			\While{$p_s^{best} \leq p_b^{best}$}
				\If{$t_b^{best} \leq t_s^{best}$}
					\State trade min($q_b^{best}, q_s^{best}$) at $p_b^{best}$
				\Else
					\State trade min($q_b^{best}, q_s^{best}$) at $p_s^{best}$
				\EndIf
				\State subtract traded quantity from $o_b^{best}$ and $o_s^{best}$
				\State remove $o_b^{best}$ and/or $o_s^{best}$ from book if filled 
				\State update trader bidding strategies
			\EndWhile
			\State update trader bidding strategies
		\EndWhile

	\end{algorithmic}
\end{algorithm}

\begin{algorithm}
\caption{ZIP Trader Bid Update Process}
\label{alg2}
	\begin{algorithmic}[1]
		\For{trader \textbf{in} market during time slot $t$}
			\LineComment{Update buyer profit margins}
			\If{trader is buyer}
				\If{$o_{last}$ was matched at price $p_{trade}$}
					\If{$p_b \geq p_{trade}$}
						\State decrease bid price $p_b$
	
					\ElsIf{$o_{last}$ is sell order \textbf{and} $p_b \leq p_{trade}$}
						\State increase bid price $p_b$ \textbf{if} $p_b \leq p_{trade}$
					\EndIf
				\ElsIf{$o_{last}$ is buy order \textbf{and} $p_b \leq p_{trade}$}
					\State increase bid price $p_b$
				\EndIf
			\LineComment{Update seller profit margins}
			\Else
				\If{$o_{last}$ was matched at price $p_{trade}$}
					\If{$p_s \leq p_{trade}$}
						\State increase bid price $p_s$
					\ElsIf{$o_{last}$ is buy order \textbf{and} $p_s \geq p_{trade}$}
						\State decrease ask price $p_s$
					\EndIf
				\ElsIf{$o_{last}$ is sell order \textbf{and} $p_s \geq p_{trade}$}
					\State decrease ask price $p_s$
				\EndIf
			\EndIf			
		\EndFor

	\end{algorithmic}
\end{algorithm}

\begin{figure*}[!t]
	\centering
	\includegraphics[width=\textwidth, keepaspectratio]{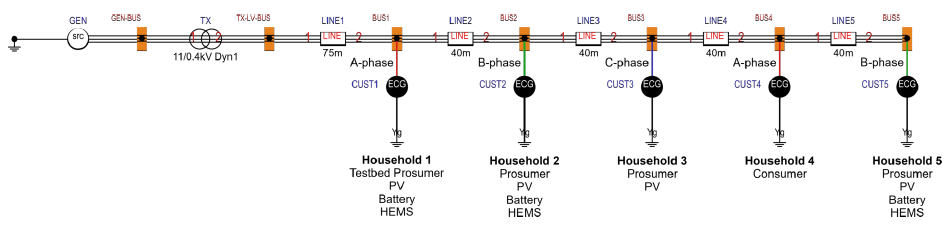}
	\caption{Simulated network model used for initial PHIL testing.}
	\label{network_model}
\end{figure*}

A simple low-voltage distribution line served as the simulated network model (Figure \ref{network_model}). The line was connected through an ideal 11/0.4kV Dyn1 distribution transformer to the wider network. Line conductors were assumed to be 7/4.50 single-core aerial bare aluminum, with impedances derived from \cite{OverheadLineStandard}. The five single-phase market participants used in the simulation were connected to the network as shown in Figure \ref{network_model}. Connections to each household were staggered over the three phases to reflect real-world residential connection standards \cite{NSWInstallationRules}. Line impedances separating each model element are shown in Table \ref{model_impedances}.

\begin{table}[!t]
	\renewcommand{\arraystretch}{1.3}
	\caption{Impedances of distribution line segments}
	\label{model_impedances}
	\centering
	
	\begin{tabular}{ ccc } 
		\hline
		Line Segment & Distance (m) & Impedance ($\Omega$) \\
		\hline 
		Line 1 & 75 & 0.0239 + 0.0218j \\ 
		Line 2 & 40 & 0.0128 + 0.0116j \\ 
		Line 3 & 40 & 0.0128 + 0.0116j \\ 
		Line 4 & 40 & 0.0128 + 0.0116j \\ 
		Line 5 & 40 & 0.0128 + 0.0116j \\
		\hline
	\end{tabular}
\end{table}

\begin{table}[!t]
	\renewcommand{\arraystretch}{1}
	\caption{Battery and PV capacities by household}
	\label{household_specs}
	\centering
	
	\begin{tabular}{ p{0.12\linewidth} p{0.12\linewidth} p{0.22\linewidth} p{0.17\linewidth} p{0.12\linewidth} } 
		\hline
		Household Number & Hardware Model & Battery Capacity (kWh) & PV Capacity (kW) & HEMS \\
		\hline 
		1 & Yes & 7.5 & 3 & Yes \\ 
		2 & No & 7.5 & 5 & Yes \\ 
		3 & No & None & 5 & No \\ 
		4 & No & None & None & No \\ 
		5 & No & 7.5  & 5 & Yes \\
		\hline
	\end{tabular}
\end{table}

Three of the households had on-site PV generation, battery storage and a home energy management system (HEMS). The HEMS was modeled as a simple linear programming (LP) optimization problem with an objective function aimed at maximizing the savings of the prosumer. The forecasting of demand and PV generation at the beginning of the simulated period was assumed to be deterministic for the sake of simplicity. The effect of forecasting errors was, therefore, outside the scope of this study. The expected income from energy trading was also not included in the HEMS, with only retail pricing being used to form the objective cost function. 

Of the two remaining households, one had only on-site PV generation and the other was a regular electricity consumer. The chosen battery and PV capacities for each household are shown in Table \ref{household_specs}.

The P2P market was operated over a one day period to verify its performance. Each prosumer was allocated an hourly PV generation profile and load profile to align with the chosen trading period duration. The profiles were adapted from actual meter data from Australian residences published by a large Australian distribution network operator (Ausgrid) \cite{AusgridData}.

\subsection{P2P Trading Results}

\begin{figure}[!t]
	\centering
	\subfloat[]{\includegraphics[width=\columnwidth, keepaspectratio]{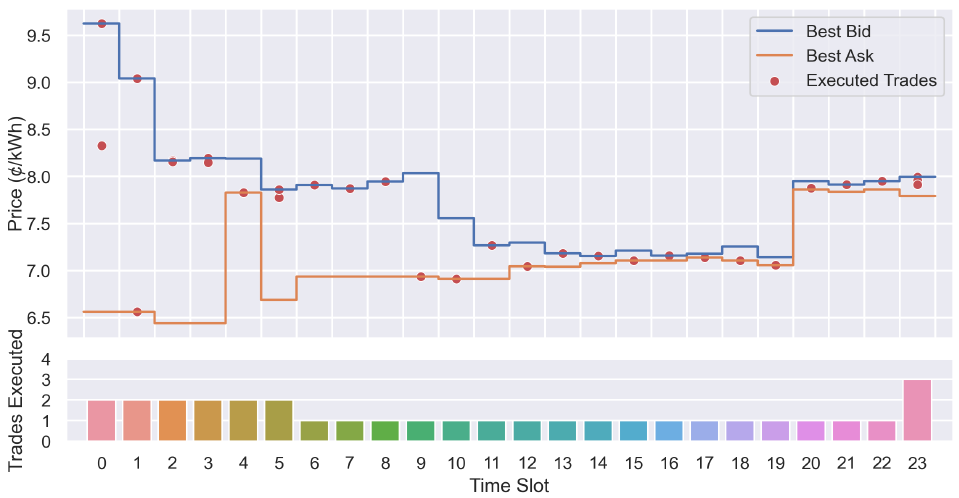} \label{exec_trades}} \\
	\subfloat[]{\includegraphics[width=\columnwidth, keepaspectratio]{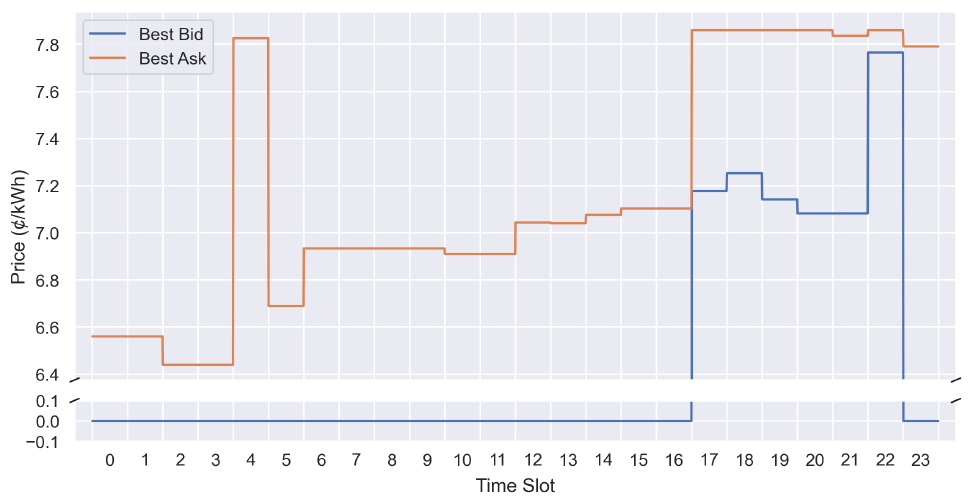} \label{unmatched}} \\
	\subfloat[]{\includegraphics[width=\columnwidth, keepaspectratio]{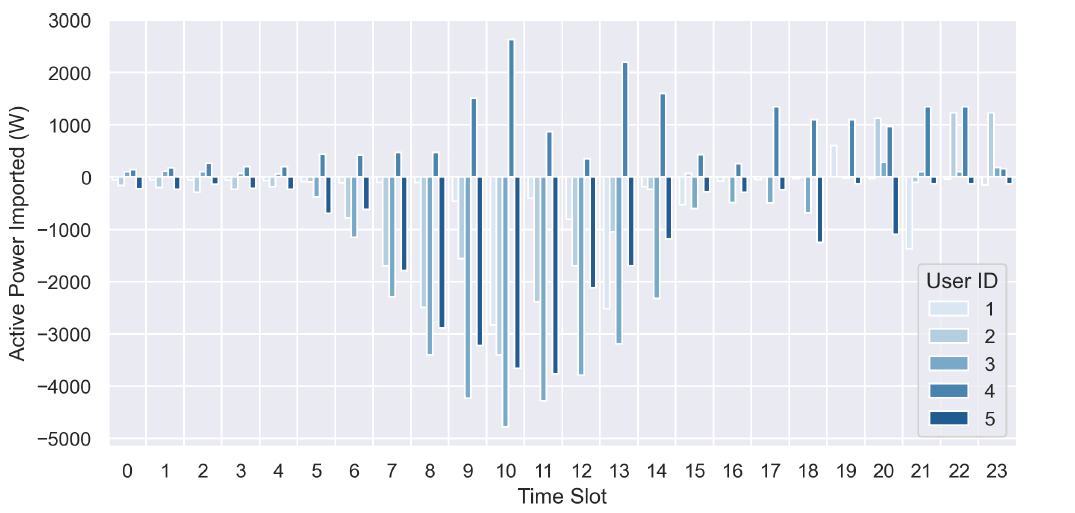} \label{grid_exchanges}}
	\caption{\ref{exec_trades}) best bid and ask prices during each trading period, \ref{unmatched}) best unmatched ask and bid following period, and \ref{grid_exchanges}) power imported by each participant.}
	\label{sim_results}
\end{figure}

The trading activity resulting from the simulation is shown in Figure \ref{sim_results}. The total value gained over a traditional retail arrangement for each household is shown in Table \ref{value_captured_results}.

\begin{figure}[!t]
	\centering
	\includegraphics[width=2.5in, keepaspectratio]{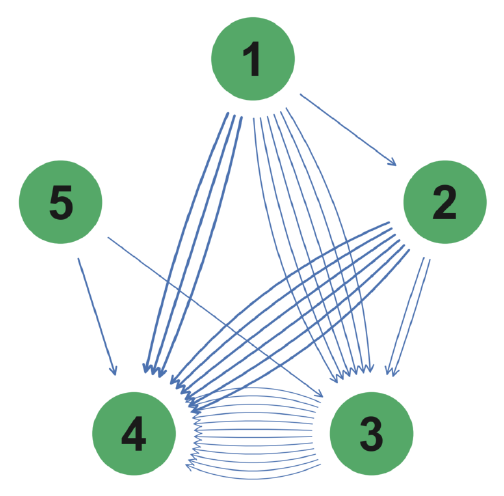}
	\caption{Trades executed between households over the simulation period.}
	\label{executed_trades}
\end{figure}

\begin{table}[!t]
	\renewcommand{\arraystretch}{1}
	\caption{Value captured by each market participant}
	\label{value_captured_results}
	\centering
	
	\begin{tabular}{ p{0.45\linewidth} p{0.08\linewidth} p{0.08\linewidth} p{0.08\linewidth} } 
		\hline
		 & Buyer & Seller & Total \\
		\hline 
		Daily Value (\$) & 2.62 & 0.19 & 2.81 \\ 
		Expected Annual Value (\$) & 958.07 & 68.41 & 1026.48 \\ 
		Expected Annual Value per Household (\$) & 191.61 & 13.68 & 205.30 \\ 
		Proportion of Value Captured (\%) & 93.34 & 6.66 & 100.00 \\ 
		\hline
	\end{tabular}
\end{table}

Figure \ref{grid_exchanges} shows that the total exports far exceeded imports over the course of the simulation. This signifies that a surplus of power generation was present during most periods, particularly during the daytime due to abundant PV generation. This caused a large number of trades to occur from prosumer households (Households 1, 2, 3 and 5) selling excess generation to the only available energy consumer (Household 4). Household 3 became available as a buyer outside of daylight hours due its the lack of on-site energy storage.

Figure \ref{exec_trades} shows that all trades were executed at prices under 10c/kWh. This was despite a FiT of 6.1c/kWh and a peak ToU price of 49.24c/kWh. The low execution prices were attributed to the large surplus of energy supply driving lower prices in the competitive market environment. The majority of value was therefore unlocked by buyers of electricity, as demonstrated in Table \ref{value_captured_results}.

Figure \ref{unmatched} reveals that potential buyers often chose to purchase power at the retail ToU rate instead of the best available ask price. This always resulted in a higher price being paid for electricity. The irrational decision-making shown here highlights a key shortcoming of the chosen ZIP trader algorithm: buyers do not account for the inelastic nature of electricity demand. Most markets (such as stock markets) allow traders the freedom to forego transactions without incurring a direct loss at the end of a trade period. In contrast, P2P electricity markets require each participant to secure enough supply to meet demand during all trading intervals. This leads to traders being forced to purchase shortages of electricity at a higher retail price if not acquired elsewhere.  

\subsection{PHIL Testing Behaviour}
\label{PHIL_behaviour}

The hardware prosumer model was allocated to Household 1 in the simulated scenario. The power interface was initially designed to transfer only the voltage at the grid connection point from the simulated network model to the hardware environment. The current measured at the amplifier output was then communicated back to the network model to complete the feedback loop, creating an arrangement similar to that shown in Figure \ref{PHIL_Interface}. The communication of the signals was done through a basic analogue connection. 

\begin{figure}[!t]
	\centering
	\subfloat[]{\includegraphics[width=\columnwidth, keepaspectratio]{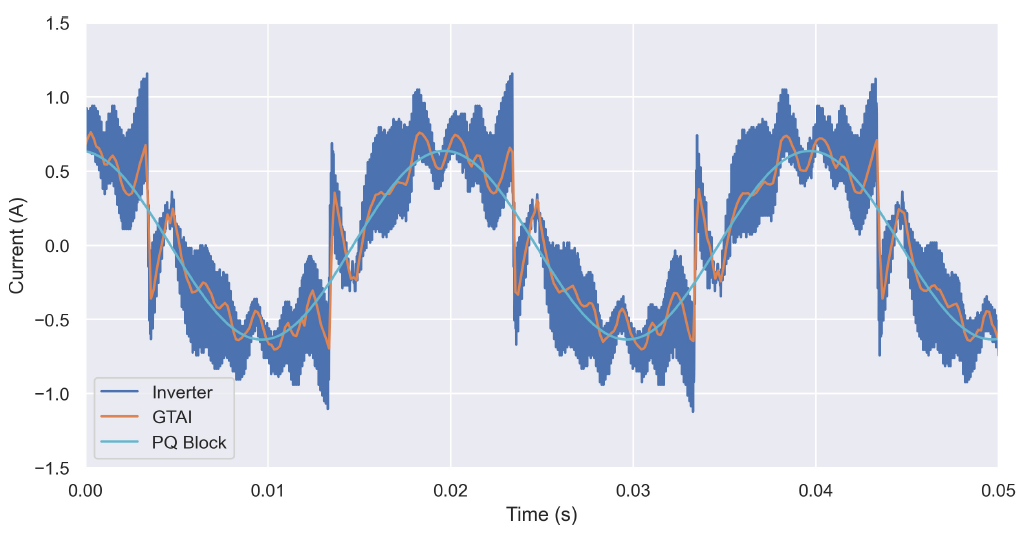} \label{pv_distortion}} \\
	\subfloat[]{\includegraphics[width=\columnwidth, keepaspectratio]{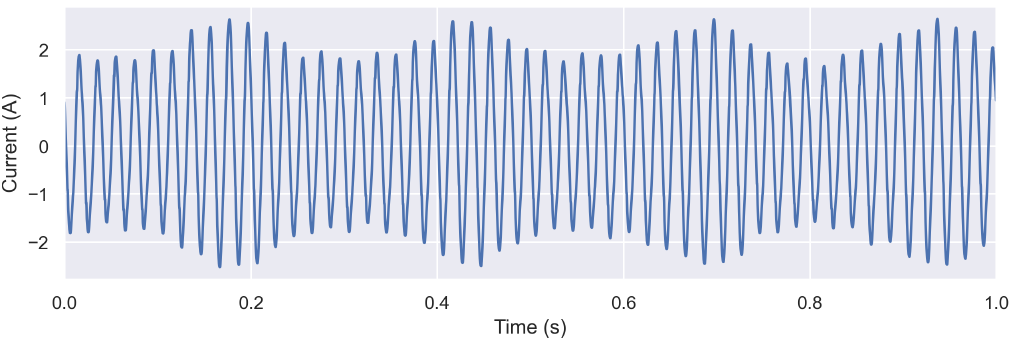} \label{battery_distortion}} \\
	\caption{\ref{pv_distortion}) comparison of Sunny Boy PV inverter distortion with the corrected waveform using the RSCAD PQ block, and \ref{battery_distortion}) current oscillations introduced by the Sunny Island battery inverter.}
	\label{stability_results}
\end{figure}

The simulation was found to be unstable under this arrangement. This was largely attributed to the high harmonic distortion present in the outputs of the commercial PV and battery inverters (Figure \ref{stability_results}). It was concluded that reductions to the total time delay of the PHIL feedback loop would be needed to properly capture this behavior. The use of shorter simulation time steps and fiber communication over analogue connections are being explored as potential solutions.

Stability was obtained in the meantime by processing the measured hardware signals in RSCAD. To do this, voltage and current were measured synchronously from the power amplifier and communicated to the simulation environment. These signals were then converted artificially into pure active and reactive power values using the in-built PQ meter block. The effect of this transformation on removing harmonic distortion can be observed in Figure \ref{pv_distortion}.

RSCAD uses equations (\ref{active_power}) and (\ref{reactive_power}) to do this operation:

\begin{equation}
P = \frac{1}{T}\int^T_{t-T} V(\omega t) I(\omega t) dt
\label{active_power}
\end{equation} 

\begin{equation}
Q = \frac{1}{T}\int^T_{t-T} V(\omega t) I(\omega t - \frac{\pi}{2}) dt
\label{reactive_power}
\end{equation} 

This method enables waveform distortion to be factored into the result by explicitly handling non-sinusoidal conditions, as discussed in \cite{nonsinepower}.

This is in contrast to calculating power values in (\ref{active_power_sine}) and (\ref{reactive_power_sine}), as done on the human-machine interface (HMI) of the power amplifier:

\begin{equation}
P = VI\cos{\theta}
\label{active_power_sine}
\end{equation} 
\begin{equation}
Q = VI\sin{\theta}
\label{reactive_power_sine}
\end{equation} 

These assume that no distortion is present by only considering the fundamental frequency of the waveform. It was discovered that this yielded values considerably different to real-time integration when applied to the measured signals. The non-sinusoidal conditions created by inverter devices can therefore have a large impact on meter device readings depending on the calculation method. This could have real-world implications in networks with a high penetration of DER.

\begin{figure}[!t]
	\centering
	\includegraphics[width=\columnwidth, keepaspectratio]{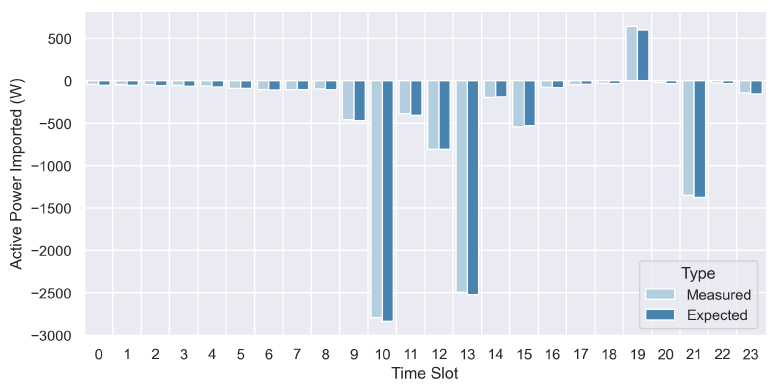}
	\caption{Measured vs expected active power injection from the hardware prosumer model.}
	\label{power_accuracy}
\end{figure}

The PHIL system maintained stability under a wide range of conditions using the PQ block conversion. Active power settings issued to the battery inverter in response to P2P trading closely matched values measured in the network model, as shown in Figure \ref{power_accuracy}. However, discrepancies could still be observed between the power traded in the P2P market and the amount actually dispatched. There is a need to further investigate methods of resolving these instances within the market framework.

\section{Conclusion}
\label{conclusion}

P2P energy trading is a viable alternative for coordinating DER. It incentivizes prosumers to preserve the local supply-demand balance through the underlying market mechanism. The competition created between buyers and sellers fosters more efficient pricing and allocations of resources when compared to traditional approaches.

Challenges are introduced through the underlying interactions between physical hardware devices. These included the high levels of harmonic distortion introduced by commercial PV and battery inverters and mismatches between traded energy and amounts actually dispatched. The non-sinusoidal conditions created by DER also highlight the importance of the method chosen for measuring real and reactive power. All of these issues are particular to the hardware used and are not easily replicable in pure software. This highlights the need for PHIL testing in the study of DER coordination strategies.

Capturing the harmonic distortion from inverter devices presented a challenge for the PHIL test environment. It was discovered that a highly-ideal power interface is necessary to preserve system stability when capturing this behavior. Reducing time delays through improved communication methods and smaller simulation time steps are being explored towards this end. The level of distortion observed could also have real-world ramifications for distribution networks and warrants additional investigation.

The effects of wireless communication methods on the scalability of P2P energy trading are yet to be investigated. Future studies will employ the RPi devices to explore the impact of data loss and communication delays. This will provide insight into the feasibility of large-scale P2P applications.

Further stability analysis is required to establish the exact operating limits of the PHIL test environment. Quantifying stability margins is critical for gauging which power system behaviors can be studied. The information can also be used to make informed decisions on future PHIL designs.

Finally, the inelastic nature of electricity demand was found to be problematic when modeling prosumer bidding strategies. ZIP traders were discovered to display irrational behavior when used in an energy market setting. Opportunities to purchase electricity in the CDA were routinely declined in favor of more expensive retail tariffs, leading to lower economic benefits for all parties. Future work is needed to refine the simulation of prosumers in P2P markets. 

\bibliographystyle{IEEEtran}
\bibliography{PHIL_Testing_of_a_Peer-to-Peer_Energy_Trading_Framework}

\end{document}